\documentstyle[prl,multicol,aps,epsf]{revtex}

\begin{document}
\draft
\input BoxedEPS
\SetepsfEPSFSpecial
\HideDisplacementBoxes

\title{Transparent Anomalous Dispersion and Superluminal Light Pulse Propagation
at a Negative Group Velocity}

\author{A. Dogariu, A. Kuzmich, and L. J. Wang}
\address{NEC Research Institute, 4 Independence Way, Princeton, NJ 08540, USA\\}
\date{September 27, 2000)\\
(Phys. Rev. A, to be published}
\maketitle

\begin{abstract}

Anomalous dispersion cannot occur in a transparent passive medium where electromagnetic 
radiation is being absorbed at all frequencies, as pointed out by Landau and Lifshitz. Here
we show, both theoretically and experimentally, that transparent linear anomalous dispersion can
occur when a gain doublet is present. Therefore, a superluminal light pulse propagation can be
observed even at a negative group velocity through a transparent medium with almost
no pulse distortion. Consequently, a {\it negative transit time} is experimentally
observed resulting in the 
peak of the incident light pulse to exit the medium even before entering it. This
counterintuitive effect is a direct result of the {\it rephasing} process owing to 
the wave nature of light and is not at odds with
either causality or Einstein's theory of special relativity.

\end{abstract}
\vspace*{.3 in}
\begin{multicols}{2}
\narrowtext

\section{Introduction}
In a seminal paper, Lord Rayleigh remarked that a pulse of light travels at the ``group velocity" 
instead of the phase velocity inside a medium \cite{rayleigh1}. In subsequent papers, 
Lord Rayleigh further  developed the
theory on transparency and opacity, and the theory of anomalous dispersion \cite{rayleigh2,rayleigh3}.
Anomalous dispersion was first studied for mechanical oscillators [3] and was later applied by Sommerfeld and Brillouin
\cite{brillouin} to light propagating in
absorptive opaque materials. They showed theoretically that inside an absorption line,
the dispersion is anomalous, resulting in a group velocity faster than $c$, the vacuum speed of light.
Such an anomalous velocity appears owing to the wave nature of light \cite{born,landau}. .

This however raised difficulties with the theory of relativity which states that no velocity can be
higher than $c$, the velocity of light in a vacuum. Brillouin pointed out \cite{brillouin} ``Group velocity,
as originally defined, became larger than $c$ or even negative within an absorption band. Such a
contradiction had to be resolved and was extensively discussed in many meetings about 1910."

In order to reconcile the superluminal group velocity in an anomalous dispersion medium
with the implied limitation from relativity,  Sommerfeld and
Brillouin pointed out that causality only requires the speed of a signal carried by light be limited by $c$,
rather than the light pulse itself which travels at the group velocity \cite{brillouin}. They further pointed out
that the correct definition of the speed of a light signal should be defined as the ``frontal velocity," 
instead of the somewhat misleadingly named ``signal velocity" defined as the speed of the half-point
of the front edge of a light pulse. A ``frontal
velocity" marks the velocity of an infinitely sharp step-function-like change in the light intensity, albeit 
even in principle such an infinitely sharp change in time requires infinite bandwidth and thus becomes impractical.
It was shown that since the infinitely sharp ``front" contains infinitely large bandwidth, a good portion
of the power would exceed any practical resonant and non-resonant (plasma) frequencies to have an effective
refractive index of unity. Thus this portion of the power will propagate exactly at $c$ to exit the
medium at the earliest time and form ``precursors" \cite{brillouin}. It was an ingenious argument that resolved 
the apparent conflict of special relativity and superluminal group velocities. 

More recently, it has been considered 
\cite{chiao1,garrett,chu,segard,aku,basov,yariv,lamb,picholle,tajima,chiao2,chiao3,tajima2,chiao4,chiao5,chiao6} 
to use various schemes including the anomalous dispersion near an absorption line\cite{garrett,chu,segard,aku},
nonlinear\cite{basov} and linear gain lines \cite{yariv,lamb,picholle,tajima,chiao2,chiao3,chiao4,chiao5,chiao6},
active plasma medium\cite{tajima,tajima2}, 
or in a tunneling barrier\cite{chiao7,chiao8} to observe these effects.
Sometimes it is applied\cite{chu,segard,aku,chiao7,chiao8}
to observe superluminal propagation of light pulses. 
Inside an absorption line, it was shown that a light pulse propagates at group velocities faster than
$c$ and can become negative with dramatic values of $-c/23,000$ \cite{chu,segard,aku,budker}.
However, in all experiments light pulses experienced either very large
absorption\cite{chu,segard} or severe reshaping \cite{chiao7} 
that sometimes resulted in controversies over the interpretations. 

In a series of papers\cite{chiao1,chiao2,chiao3,chiao4,chiao5,chiao6,chiao7,chiao8}, Chiao and coworkers
showed theoretically that anomalous dispersion can occur inside a transparent material, particularly on
the DC side of a single gain line. It was predicted that by using a gain doublet\cite{chiao4}, it is possible to
obtain a transparent anomalous dispersion region where the group velocity of a light pulse exceeds $c$
with almost no pulse distortion. 

Here we use gain-assisted linear anomalous dispersion to demonstrate superluminal
light pulse propagation with a negative group velocity through a transparent atomic medium \cite{wang}. 
We place two Raman gain peaks closely to obtain an essentially lossless anomalous dispersion region that results
in a superluminal group velocity. 
The group velocity of a pulse in this region exceeds c and can even become negative\cite{chiao4,chiao5},
while the shape of the pulse is preserved. We measured a negative group velocity index of $n_{g}= -315(\pm5)$. 
Experimentally, a light pulse propagating through the atomic vapor cell exits from it 
earlier than propagating through the same distance in vacuum by a time difference 
that is 315 times of the vacuum light propagation time $L/c=0.2$ ns.  Thus, the peak of the pulse exits
the cell before it even enters. This counterintuitive effect is a consequence of the wave nature of light
and can be well explained invoking the rephasing process in an anomalously dispersive medium.
The observed superluminal light pulse propagation is not at odds with causality, being a direct consequence
of classical interference between its different frequency components in an anomalous dispersion region.

\section{Theory of Transparent Anomalous Dispersion}

For all transparent matter at a thermal equilibrium such that it is absorptive at
all frequencies in the electromagnetic spectrum, the medium's optical dispersion
is normal \cite{landau}. In other words, for transparent media, 
the optical refractive index always
increases when the frequency of the optical excitation increases.
Particularly, Landau and Lifshitz showed that under the 
condition
\begin{equation}
Im \,[\chi(\nu)] \ge 0, \,\,\, \makebox{for any}\,\, \nu 
\end{equation}
and in the special case for media with a magnetic permeability $\mu (\nu)=1$, 
two inequalities hold simultaneously:
\begin{eqnarray}
n_{g}(\nu)&=&\frac{d\,[n(\nu)\nu]}{d\nu} > n(\nu),\hspace*{.25 in} \nonumber \\
\hbox{and} \hspace*{.25 in} 
n_{g}(\nu)&=&\frac{d\,[n(\nu)\nu]}{d\nu} > \frac{1}{n(\nu)}.
\end{eqnarray}
Here $n_{g}$ is the group velocity index: $v_{g}=c/n_{g}$, where $v_{g}$ is the group velocity.

In the case of dielectric media, when an incident light wave's frequency is below the
bandgap where the media are transparent, the refractive index $n(\nu)> 1$. Hence,
the group index $n_{g}> n> 1,$ resulting in normal dispersion and the group velocity
of light pulses slower than $c$. On the other hand, in the case of metals when the incident
light's frequency is higher than the plasma frequency $\nu_{P}$, the metal becomes transparent
with a refractive index $n(\nu)< 1$. In this case, the second inequality in Eq. (2) becomes
more strict and results in $n_{g}> 1/n > 1$. Therefore, the group velocity of a light pulse 
propagating through metal is also slower than $c$, the vacuum speed of light. 

These conditions are direct results of the Kramers-Kronig relations. Hence, for
media in a passive state under Eq.(1) where electromagnetic irradiation at all
frequencies are transformed into heat and subsequently dissipated, 
anomalous dispersion and transparency cannot occur simultaneously 
in the same frequency region. 

However, for media with gain, the general assumption in Eq.(1) no longer holds.
In a gain medium such as that of an electronic amplifier (e.g. an operational amplifier)
or an optical
amplifier commonly used for lasers, the imaginary part of the electric susceptibility $\chi(\nu)$
can become negative over a narrow frequency region where gain occurs. Hence, the general
results of the inequalities (2) no longer apply and anomalous dispersion can
happen in a transparent media, resulting in superluminal group velocities that can even
become negative.

\subsection{Classical Theory of Negative Group Velocity}

In a transparent medium where absorption is negligible,
the optical dispersion is primarily governed by
its refractive index and the dispersion relation can be written:
\begin{equation}
k(\nu)=2\pi\frac{n(\nu)\cdot\nu}{c}, 
\end{equation}
where $k(\nu)$ is the wave number for the wave component of frequency $\nu$. 
The group velocity is hence given by
\begin{equation}
v_{g}=\frac{c}{n_{g}}=Re [\frac{d\omega}{d k}]=\frac{c}{Re[n+\nu d n/d\nu]}
\approx \frac{c}{n+\nu d n/d\nu}, 
\end{equation}
where 
\begin{equation}
n_{g}=n+\nu\, \frac{d n}{d\nu} 
\end{equation}
is the group velocity index. In all transparent media where dispersion is normal,
we have $dn/d\nu \ge 0$, resulting in $v_{g}\le c$. In special cases where 
this normal dispersion is very steep over a narrow frequency region such as in the
case of Electromagnetically Induced Transparency (EIT)\cite{harris,scully} that $\nu dn/d\nu \gg 1$, the
group velocity can be reduced to as slow as $8 m\cdot sec^{-1}$ \cite{xiao,hau,welch,budker}

Conversely, in a region of transparent anomalous dispersion where $\nu dn/d\nu \ll -1$, 
a negative group velocity index is obtained. If the index of refraction decreases
rapidly enough with frequency, the group velocity becomes negative. Of course,
it is well known that inside an absorption line, the refractive index takes a
steep drop \cite{brillouin} resulting in an ``anomalous dispersion" and consequently
a negative group velocity \cite{chu}. However, the
associated heavy absorption which results in opacity makes it difficult to study
these effects.

The earlier work of  Chiao and coworkers 
\cite{chiao1,chiao2,chiao4} predicted that it is possible to study steep anomalous dispersion
and negative group velocity in a transparent medium where gain rather than absorption
occurs at the interesting frequency regions. It is important to have a transparent
medium where the majority of the electromagnetic radiation energy is kept in the
light fields rather than being dissipated inside the medium via absorption.

Let us start by considering a classical Lorentz oscillator model of the refractive index.
The electric displacement is given by:
\begin{equation}
D=\varepsilon_{0}\, E+P=\varepsilon_{0} (1+\chi) E =\varepsilon_{0}\, E(1+ N\alpha),
\end{equation}
where $N$ is the atomic density and $\alpha$ is the atomic polarizability.
The polarization density $P=-\varepsilon_{0}Ne\,x=\varepsilon_{0} N\alpha E$ can be obtained 
using a simple Lorentz model.

In order to obtain the dipole polarization  $p=-e\,x$ for a bound charge with an intrinsic 
angular frequency $\omega_{0}=2\pi\nu_{0}$ and an angular damping rate $\Gamma=4\pi\gamma$, 
we start from the equation of motion of the electron:
\begin{equation}
\ddot{x}+\Gamma\dot{x}+\omega_{0}^{2}x=-\frac{eE}{m}\,=-\frac{eE_{0}}{m}\, e^{-i\omega t}.
\end{equation}
Hence, we obtain,
\begin{equation}
x=\frac{eE}{m}\,\frac{1}{\omega^{2}-\omega_{0}^{2}+i\omega\Gamma}
\approx \frac{eE}{2m\omega_{0}}\,\frac{1}{\omega-\omega_{0}+i\Gamma/2},
\end{equation}
where the approximation is good as long as $\omega_{0}\gg\Gamma$. 
We further obtain for the polarizability,
\begin{equation}
\alpha = -\frac{e^2}{2m\omega_{0}}\,\frac{1}{\omega-\omega_{0}+i\Gamma/2}
= -\frac{e^2}{4\pi m\omega_{0}}\times\,\frac{1}{\nu-\nu_{0}+i\gamma}.
\end{equation}
The dielectric susceptibility of the medium thus can be written:
\begin{equation}
\chi(\nu)= -\frac{Ne^2}{4\pi\varepsilon_{0} m\omega_{0}}\times\,\frac{1}{\nu-\nu_{0}+i\gamma}
= -f\times\,\frac{M}{\nu-\nu_{0}+i\gamma},
\end{equation}
where $M=\nu_{P}^{2}/\nu_{0}$ with $\nu_{P}$ being the effective plasma frequency and $f$
being the oscillator strength. When two absorption lines are placed nearby with equal oscillator
strengths $f_{1}=f_{2}=1$, the dielectric susceptibility can be written:
\begin{equation}
\chi(\nu)= -\,\frac{M}{\nu-\nu_{1}+i\gamma}-\,\frac{M}{\nu-\nu_{2}+i\gamma}.
\end{equation} 
For a narrow frequency region in the middle between the two absorption lines, a steep normal
dispersion region occurs resulting in an ultra-slow group velocity \cite{hau,welch,budker}.

Conversely, for gain lines, a negative oscillator strength $f=-1$ is assigned \cite{chiao2}. 
Hence between
two closely placed gain lines, the effective dielectric constant can be obtained:
\begin{equation}
\epsilon(\nu)=1+\chi(\nu)
=1+ \frac{M}{\nu-\nu_{1}+i\gamma}+\,\frac{M}{\nu-\nu_{2}+i\gamma}.
\end{equation} 
For dilute gaseous medium, we obtain from Eq.(12) for the refractive index
$n(\nu)=n'(\nu)+i\,n''(\nu)=1+\chi(\nu)/2$ and the real and imaginary parts of the
refractive index are plotted in Fig.1. It is evident from Fig.1 that a steep
anomalous dispersion region appears without the heavy absorption present. In fact,
a residual gain persists. Furthermore, with the correct choice of experimental 
parameters, the steep drop of refractive index as a function of frequency can
be made a mostly linear one in this region. Thus a light pulse with a frequency bandwidth within
this narrow linear anomalous dispersion region will experience almost no change
in pulse shape upon propagating through such a medium.

\begin{figure}
\BoxedEPSF{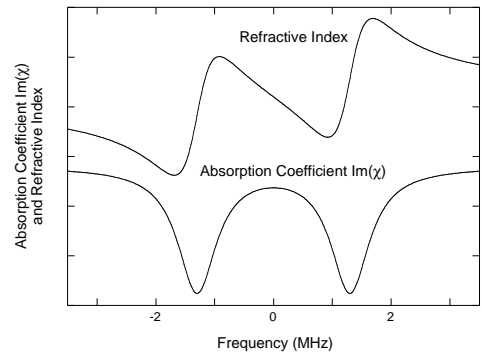 scaled 400}
\caption{Gain-assisted anomalous dispersion. Figure shows frequency-dependent gain coefficient and
refractive index.}
\end{figure}
Now, we consider the propagation of a light pulse of an arbitrary shape through a
transparent anomalous dispersing medium of a length $L$ as illustrated in Fig.2.
For a scalar light pulse that is decomposed into its positive and negative frequency parts:
\begin{equation}
E(z,t)=E^{(+)}(z,t)+E^{(-)}(z,t),
\end{equation} 
we have for its Fourier decomposition:
\begin{equation}
E^{(+)}(z,t)=\frac{1}{\sqrt{2\pi}}\int d\omega \tilde{E}^{(+)}(\omega)\, e^{-i\,(\omega t-k\cdot z)}.
\end{equation} 
At the entrance of the medium where we denote by $z=0$, we can rewrite the above 
expression as 
\begin{eqnarray}
E^{(+)}(0,t)&=&\frac{1}{\sqrt{2\pi}} e^{-i\,\omega_{0}t}\int d(\omega-\omega_{0})
\tilde{E}^{(+)}(\omega-\omega_{0}) \nonumber\\ 
&\times&e^{-i\,(\omega-\omega_{0})t},
\end{eqnarray} 
where $\omega_{0}$ is the carrier frequency of the light pulse. Inside the transparent
anomalous dispersion medium, if over the narrow bandwidth of the incident light pulse
$E(\omega-\omega_{0})$, the gain is essentially unity, the propagation is 
governed by the wave vector $k(\omega)$. Using a Taylor series expansion of the 
wave vector:
\begin{equation}
k(\omega)=k(\omega_{0})+\frac{1}{v_{g}}(\omega-\omega_{0})
+\frac{1}{2}\left. \frac{d^{2}k}{d\omega^{2}}\right|_{\omega_{0}} 
\cdot(\omega-\omega_{0})^{2},
\end{equation} 
where the expansion is carried to the quadratic term.
Nonlinear terms in the expansion of Eq.(16) are often
associated with ``Group Velocity Dispersion (GVD)," or chirping terms causing pulse distortion. 
When these nonlinear terms in
Eq.(16) are negligible, i.e., the dispersion is essentially linear, from Eq.(14) and (16) 
we obtain:
\begin{eqnarray}
E^{(+)}(L,t)&=&\frac{1}{\sqrt{2\pi}} e^{-i\,(\omega_{0}t-k_{0}\cdot L)}\int d(\omega-\omega_{0})
\tilde{E}^{(+)}(\omega-\omega_{0}) \nonumber \\ 
&\times&e^{-i\,(\omega-\omega_{0})\cdot(t-L/v_{g})}.
\end{eqnarray} 

\begin{figure}
\BoxedEPSF{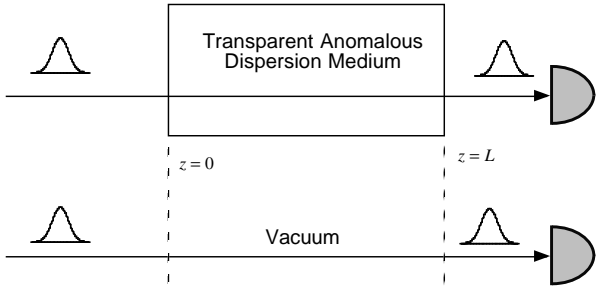 scaled 550}
\vspace*{.15 in}
\caption{Pulse propagation through a medium of a length $L$ and a group velocity
index $v_{g}=c/(n+\nu\,dn/d\nu)$. Pulse propagation through the same length in vacuum
is also shown for comparison.}
\end{figure}

Hence, the intensity of the light pulse as a function of time measured with a detector 
(shown in Fig.2), $I(L,t)$, is related to the incident pulse's time-dependent intensity by:
\begin{eqnarray}
I(L,t)&=&\frac{\varepsilon_{0}cA}{2}|E^{(+)}(L,t)|^{2}=\frac{\varepsilon_{0}cA}{2}|E(0,t-L/v_{g})|^{2} \nonumber \\
&=&I(0,t-L/v_{g}),
\end{eqnarray}
where $A$ is the beam area.
  
Ordinarily, in a normal dispersion medium, the group velocity $v_{g}< c$. Hence, the
output intensity of a pulse propagating through the medium is retarded by the propagation
time $L/v_{g}$, resulting in a delay longer than the vacuum transit time $L/c$. 
In a transparent anomalous dispersion medium,
the group velocity $v_{g}=c/[n+\nu\,dn/d\nu]$ can exceed $c$ provided the anomalous
dispersion is sufficiently strong such that: $n+\nu\,dn/d\nu < 1$. In this case, the
group velocity becomes superluminal: $v_{g} > c$, resulting in a ``{\it superluminal transit time}:"
$L/v_{g} <L/c$, the vacuum transit time. Hence the output pulse's time-varying 
profile is related to the input pulse by a delay that is shorter than the vacuum transit 
time $L/c$ resulting in a superluminal propagation of the light pulse.

Furthermore, when the transparent anomalous dispersion becomes stronger to yield:
$n+\nu\,dn/d\nu =0$, the group velocity $v_{g}=c/[n+\nu\,dn/d\nu]$ approaches
infinity, resulting in a ``{\it zero transit time}", such that Eq.(18) gives
$I(L,t)=I(0,t-L/v_{g})=I(0,t)$. In this case, the output pulse and the input pulse 
vary the same way in time and there is no time delay experienced by the
pulse propagating through the medium which has a length of $L$.

Finally, when the transparent anomalous dispersion becomes very steep such as for
the case illustrated in Fig.1, the dispersive term $\nu\,dn/d\nu$ which is negative 
becomes very large in its magnitude such that $|\nu\,dn/d\nu|\gg 1$, resulting in
a negative group velocity $v_{g}=c/[n+\nu\,dn/d\nu] < 0$. In this case, Eq.(18)
gives $I(L,t)=I(0,t+|L/v_{g}|)$, where the quantity $|L/v_{g}|=|n_{g}|\,L/c$ 
is positive and can becomes very large compared to the vacuum transit time
$L/c$. This means that the intensity at the output of the medium of length $L$,
$I(L,t)$, will vary in time {\it earlier} than that of the input pulse $I(0,t)$.
Thus in this case, a ``{\it negative transit time}" can be observed.
The time difference between the output pulse and the input pulse in the form of a
pulse advancement, is $|n_{g}|$ fold of the vacuum transit time $L/c$. Practically,
since the shape of the pulse is not changed, this results in a rather
counterintuitive phenomenon where a certain part of the light pulse 
has already exited the medium before the corresponding part of the incident
light pulse even enters by a time difference that is $|n_{g}|$ times of the
vacuum transit time $L/c$.

This rather counterintuitive effect is a result of the wave nature of light.

\subsection{Quantum Theory of Atomic Response in Transparent Anomalous Dispersion}

In order to correctly model the electromagnetic responses of a dilute atomic gaseous medium,
one must compute its dielectric susceptibility using the quantum mechanical treatment. 
In this section, we provide a simplified yet realistic model of a transparent anomalous
dispersion medium realized using two Raman gain lines in a $\Lambda$-system shown in Fig.3.

In a simplified $\Lambda$-system, two continuous wave (CW) Raman pump light
fields are present for the creation of the gain doublet shown in Fig.1. The two Raman
pump fields are different in frequency by a small amount $2\Delta\nu$. For simplicity,
we will first ignore the Doppler shift and assume that the atoms are at rest.

We begin by treating the simple case of a single Raman pump field and compute the linear
dielectric susceptibility for the Raman probe field. Let us first suppose that all atoms are 
initially prepared in an energy ground state $|1\rangle$ via
optical pumping. In the atomic system, another
ground state $|2\rangle$ is also present and a Raman transition from $|1\rangle$ to $|2\rangle$
can take place via an off-resonance two-photon transition through an excited state $|0\rangle$
by absorbing a pump photon and emitting a probe photon, with a corresponding 
transition from the state $|1\rangle$ to $|2\rangle$. 
The
effective Hamiltonian for such a $\Lambda$-system coupled with the Raman pump and probe fields 
can be written:
\begin{equation}
\hat{H}=\hat{H}_{0}+\hat{H}_{I},
\end{equation}
where
\begin{equation}
\hat{H}_{0}=-\hbar\omega_{01}\,|1\rangle\langle 1|-\hbar\omega_{02}\,|2\rangle\langle 2|,
\end{equation}
with $\omega_{0j}=(E_{0}-E_{j})/\hbar$ for $j=1,2$.
The interaction Hamiltonian can be written:
\begin{equation}
\hat{H}_{I} = -\hbar\Omega_{1}\,e^{-i\omega_{1}t}\,|0\rangle\langle 1|
-\hbar\Omega_{2}\,e^{-i\omega_{2}t}\,|0\rangle\langle 2| + H.c.,
\end{equation}
where  $\Omega_{j}=(\hat{e}_{j}\cdot\mu_{0j})\,E_{j}/2\hbar$ ($j=1,2$) is the
Rabi frequency for the pump and the probe
fields, respectively. $\mu_{0j}$ is the dipole moment between the states
$|j\rangle$ and the excited state $|0\rangle$. Because $E_{1}$ is a strong Raman pump
field while $E_{2}$ is a weak Raman probe field, therefore, we have $|\Omega_{1}|
\gg |\Omega_{2}|$.

\begin{figure}
\hspace*{.2in}
\BoxedEPSF{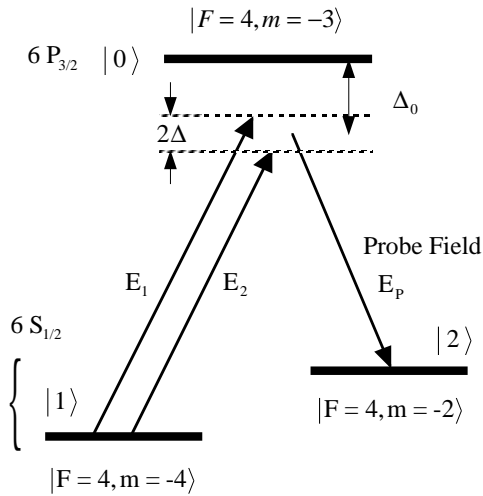 scaled 800}
\vspace*{.3 in}
\caption{Schematic atomic level diagram.}
\end{figure}

Without involving complicated density operator treatment which is only necessary
in a full quantum theory where the electromagnetic field is also quantized, we
can apply here a simplified state vector treatment in order to obtain the dielectric
susceptibility of a dilute gaseous atomic medium. The state vector of an atom can
be written:
\begin{equation}
|\psi(t)\rangle = a_{0}(t)\,|0\rangle +a_{1}(t)\,e^{i\omega_{01}t}\,|1\rangle
+a_{2}(t)\,e^{i\omega_{02}t}\,|2\rangle .
\end{equation}
Hence, we obtain for the amplitude of the excited state $|0\rangle$,
\begin{equation}
\dot{a}_{0}(t)=i\,\Omega_{1}e^{-i\,\Delta_{1}t}a_{1}
+i\,\Omega_{2}e^{-i\,\Delta_{2}t}a_{2} ,
\end{equation}
where $\Delta_{j}=\omega_{j}-\omega_{0j}$, $(j=1,2)$ represent the detuning of the Raman
pump and probe beams, respectively. To the lowest order of approximation, 
we have for the state amplitudes $a_{1}\approx 1$ and $a_{2}\approx 0$. We further
note the fact that the common detuning $\Delta_{0}=(\Delta_{1}+\Delta_{2})/2$ is
much greater than the differential detuning $\Delta_{1}-\Delta_{2}$. Under these
conditions, we have from Eq.(23):
\begin{equation}
{a}_{0}(t)\approx -\frac{\Omega_{1}}{\Delta_{0}}e^{-i\,\Delta_{1}t}a_{1} .
\end{equation}
Note here that we have omitted the decay of the excited state amplitude $a_{0}$ in Eq.(23).
First, the broadened line width is still much narrower than the common detuning $\Delta_{0}$
such that Eq.(24) is a good approximation. Second, in a Raman scheme as used in the
experiment, the major factor of the Raman transition broadening is due to transit time
broadening rather than the effective excited state decay rate: 
$(\Omega_{1}^{2}/\Delta_{0}^{2})/T_{2}$, where $T_{2}$ is the excited state lifetime.
Hence, Eq.(24) is a good approximation.

Next, we obtain the equation of motion for the amplitude of the Raman final state 
$|2\rangle$:
\begin{equation}
\dot{a}_{2}(t)=-i\,\frac{\Omega_{1}\Omega_{2}^{*}}{\Delta_{0}}
e^{-i\,(\Delta_{1}-\Delta_{2})\,t}a_{1}
-\Gamma \,a_{2} .
\end{equation}
Here, a phenomenological decay rate $\Gamma$ is used to account for the Raman transition
line broadening. Solving Eq.(25), we obtain
\begin{equation}
{a}_{2}(t)=\left(\frac{\Omega_{1}\Omega_{2}^{*}}{\Delta_{0}}\right)\cdot
\frac{1}{(\Delta_{1}-\Delta_{2})+i\,\Gamma}
e^{-i\,(\Delta_{1}-\Delta_{2})\,t}a_{1}.
\end{equation}

Using Eq.(24) and Eq.(26) and the definition of the dielectric polarization
$P=N\,\mu_{2,0}\,\rho_{02}=\chi\varepsilon_{0}\,E_{2}$ where the density matrix element
$\rho_{02}=a_{0}\,a_{2}^{*}\,e^{-i\,\omega_{02}t}$, we obtain for the dielectric
susceptibility for the Raman probe field $E_{2}$,
\begin{equation}
\chi(\Delta_{2})=-N\cdot \frac{|\mu_{02}\cdot\hat{e}_{2}|^{2}}
{2\hbar\,\varepsilon_{0}}\cdot
\frac{|\Omega_{1}|^{2}}{\Delta_{0}^{2}}\cdot
\frac{1}{(\Delta_{1}-\Delta_{2})-i\,\Gamma} .
\end{equation}
Using the fact that $\Delta_{1}-\Delta_{2}=-2\pi\,[\nu_{2}-(\nu_{1}-\nu_{01}+\nu_{02})]$,
and redefining $\nu_{0}= (\nu_{1}-\nu_{01}+\nu_{02})$ while replacing $\nu_{2}$ with $\nu$,
we can rewrite Eq.(27) as:
\begin{equation}
\chi(\nu)=\frac{M}{\nu-\nu_{0}+i\,\gamma} ,
\end{equation}
where the factor
\begin{equation}
M=N\cdot \frac{|\mu_{02}\cdot\hat{e}_{2}|^{2}}
{4\pi\hbar\,\varepsilon_{0}}\cdot
\frac{|\Omega_{1}|^{2}}{\Delta_{0}^{2}}.
\end{equation}

When two Raman pump beams of equal strengths and
slightly different frequencies $\nu_{0}-\Delta\nu$ and
$\nu_{0}+\Delta\nu$ are present, the dielectric susceptibility of
the Raman probe field becomes:
\begin{equation}
\chi(\nu)=\frac{M}{(\nu-\nu_{0}-\Delta\nu)+i\,\gamma}+
\frac{M}{(\nu-\nu_{0}+\Delta\nu)+i\,\gamma} ,
\end{equation}
similar to that obtained using a classical Lorentz model given in Eq.(11).

However, in a gaseous atomic medium, there is Doppler broadening that demands the
expression in Eq.(29) and Eq.(30) being modified. Specifically, the common detuning in
Eq.(29), $\Delta_{0}$, needs to be replaced by the shifted detuning $\Delta_{0}+\nu_{0}\, V/c$,
where $\nu_{0}$ is the carrier frequency of the light field and $V$ is the velocity of those
atoms in a specific velocity group $G(V)$ for which the atoms move along the light propagation
direction with a speed $V$. Owing to the co-linear propagation geometry, the Doppler shifts
in both $\Delta_{1}$ and $\Delta_{2}$ are essentially the same. Hence,
the shift in frequencies in $\Delta_{1}-\Delta_{2}$ or
$\nu-\nu_{0}\pm\Delta\nu$ are essential zeo to within an error of 
$\Delta\nu\cdot V/c$ which is negligible. Therefore, Eq.(30) need only be modified
by replacing the expression of the $M$-factor with:
\begin{equation}
M=N\cdot\frac{|\mu_{02}\cdot\hat{e_{2}}|^{2}}
{4\pi\hbar\,\varepsilon_{0}}\,
\int dV \frac{|\Omega_{1}|^{2}}{(\Delta_{0}+\nu\cdot V/c)^{2}}\, G(V).
\end{equation}

The effects of this modification are two fold. First, the quadratic dependence on 
$1/(\Delta_{0}+\nu\cdot V/c)^{2}$ in Eq.(31) is an even function which prevents cancellation
of the effect due to Doppler broadening. This is a direct result of using Raman transitions.
Second, for some velocity groups where the shifted detuning $\Delta_{0}+\nu\cdot V/c$ may
vanish. However, for these velocity groups, the Raman pump beams act like reversed optical
pumping which empty these velocity groups such that $G(V_{0})=0$. This is similar to the
``spectral hole burning" effect commonly know in laser physics. Furthermore, the atoms
reversely pumped away in these velocity groups act like a broadband weak absorber that
helps to compensate the residual gain as shown in Fig.1.

The spectral linewidths of the gain lines are approximately given by 
\begin{equation}
\gamma\approx\frac{v_{R}}{2\pi\,w_{0}},
\end{equation}
due to transit broadening \cite{thomas}. Here $v_{R}$ is the mean atomic velocity in the radial direction and $w_{0}$
is the mean radius of the Gaussian beam.

\subsection{Parameter Dependence in Gaussian Pulse Propagation}

The primary result of Eq.(18), $I(L,t)=I(0,t-L/v_{g})$, applies to light pulses of arbitrary
shapes, provided that over the pulse bandwidths the dispersion is transparent and 
linear. In order to further examine the implications and limits of pulse propagation 
in a transparent anomalous dispersion medium, here we treat the propagation of a Gaussian
pulse.

For the propagation of a Gaussian pulse through a transparent anomalous dispersion medium,
there are three important parameters: the peak intensity, the pulse advancement, and the
pulse shape distortion. In this section, we calculate the parameter dependence of these
factors on various experimental parameters. These parameter dependence will be compared 
with experimental results.

To begin, let us refer to Fig.2 and
consider a Gaussian pulse of a temporal duration $\tau$ and an angular carrier
frequency $\omega_{0}$, for which the incident waveform becomes:
\begin{equation}
E(z,t)=E_{0}\,e^{-(t-z/c)^2/2\tau^{2}}\,e^{-i\omega_{0}\,(t-z/c)} \hspace*{.25 in} (z<0).
\end{equation}
At the entrance of a transparent anomalous dispersion medium $z=0$, the electric field
hence varies in time as:
\begin{equation}
E(0,t)=E_{0}\,e^{-t^2/2\tau^{2}}\,e^{-i\omega_{0}t} .
\end{equation}
Therefore, at the exit surface of a medium of length $L$, the electric field varies in time:
\begin{eqnarray}
E(L,t) &=&\frac{\tau E_{0}\,e^{-\eta L}}{\sqrt{2\pi}}\,\int e^{-\tau^2 (\omega-\omega_{0})^{2}/2}\,e^{-i(\omega_{0}t-k\, L)} 
d\omega \nonumber \\
&=& \alpha\, E_{0} e^{-\beta^{2}\,(t-L/v_{g})^{2}/2\tau^2}\,e^{-i\omega_{0}\,[t-n'(\omega_{0})L/c]},
\end{eqnarray}
where 
\begin{equation}
\alpha=e^{- n''(\omega_{0})\,\omega_{0}\,L/c-\eta\,L}\cdot\beta
=e^{g\,L}\cdot\beta
\end{equation}
describes the change in pulse amplitude due to residual amplification and a weak broadband 
absorption factor $\eta$, resulting in a phenomenological gain coefficient $gL$.
The change in the width of the pulse is entirely characterized by the parameter
\begin{equation}
\beta=\left[1-i\frac{L}{\tau^{2}}\,\left.\frac{d^{2}k}
{d\omega^{2}}\right|_{\omega_{0}}\right]^{-\frac{1}{2}} .
\end{equation}

Using the simple expression in Eq.(30) and (31), here we calculate the scaling
dependence of various quantities which can be compared with experimental results.

First, using Eq.(30), we obtain a residual gain coefficient of
\begin{equation}
g=-n''(\omega_{0})\,\omega_{0}/c-\eta=\frac{2\pi\,\nu_{0}\,M}{c}\cdot\frac{\gamma}
{(\Delta\nu^{2}+\gamma^{2})}-\eta.
\end{equation}
We note here that the factor $M$ is linear in the atomic density $N$ and the Raman pump
beam power $|\Omega_{1}|^{2}$ as given in Eq.(31). 
Meanwhile, the residual loss coefficient $\eta$, which is due to absorption of atoms reversely
pumped away into other atomic states is also linear in the atomic density $N$. 
Its dependence on the Raman pump power is more complex since power broadening and 
common detuning are involved.

Next, we calculate the pulse advance (negative delay), $\Delta T$,
for a Gaussian pulse propagating through an anomalous dispersion medium of length $L$. 
Referring the experimental situation illustrated in Fig.2, a light pulse propagating through
a transparent anomalous dispersion medium of a length $L$ is advanced compared to the same
pulse propagating through the same distance in a vacuum by,
\begin{equation}
\Delta T= (1-n_{g})\, \frac{L}{c} =-\nu_{0}\left.\frac{dn}{d\nu}\right|_{\nu_{0}}\,\frac{L}{c}.
\end{equation}
Using the expression in Eq.(30), we have for the pulse advance:
\begin{equation}
\Delta T= \nu_{0}\, M\,\frac{L}{c}\cdot\frac{\Delta\nu^{2}-\gamma^{2}}
{(\Delta\nu^{2}+\gamma^{2})^{2}}.
\end{equation}
The pulse propagating through the transparent anomalous dispersion medium is advanced
provided that $\Delta\nu>\gamma$.
Here we further note that the pulse advance, $\Delta T$, is linear in $M$ which is linear in the
atomic density $N$ and the Raman pump power $|\Omega_{1}|^{2}$. These parameter dependences
will be compared with experimental results.

Finally, we compute the pulse width change factor $\beta$. We notice from Eq.(37)
that $\beta$ depends on the derivative of the group velocity index $n_{g}$. 
From Fig.1, it is apparent that the frequency-dependent change 
of the real part of the refractive index $n^{'}(\nu)$ is essentially linear in the narrow
frequency region between the two gain lines. It is the imaginary part, i.e., the gain coefficient
that appears to have a quadratic change as a function of frequency.
Therefore, the majority of the potential change 
in the width of an incident pulse will be due to the quadratic change in the imaginary part
of the refractive index.  We obtain using Eq.(30),
\begin{equation}
\left.\frac{d^{2}k}{d\omega^{2}}\right|_{\omega_{0}}
=-iM\nu_{0}\frac{\gamma\,(3\Delta\nu^{2}-\gamma^{2})}{\pi c\,(\Delta\nu^{2}+\gamma^{2})^{3}}.
\end{equation}
Hence, from Eq.(7), the pulse width distortion factor $\beta$ becomes:
\begin{equation}
\beta=\left[1- M\nu_{0}\frac{L}{\pi c\,\tau^{2}}\,
\frac{\gamma\,(3\Delta\nu^{2}-\gamma^{2})}{(\Delta\nu^{2}+\gamma^{2})^{3}}\right]
^{-\frac{1}{2}} .
\end{equation}
When various parameters such as the
$M$ factor (linear in Raman pump power and atomic density) and the length of the atomic chamber $L$, 
are significantly increased, or the pulse duration $\tau$ is significantly decreased, 
the factor $\beta$ will become reasonably different from (larger than) unity. 
In these cases, $\beta>1$ will result in a narrowing in the pulse's temporal width. 

In the present experiment, this pulse distortion factor $\beta$ is only $1.002$.

\section{Experiments}

From the theoretical treatment above, we see that a variety of parameters have to be
taken into consideration in experimentation. From an experimental point of view, one
must satisfy the following requirements. First, a gain doublet must be obtained for which
the anomalous dispersion between the gain lines can become linear to avoid any pulse distortion.
In previous work, excited state population inversion was considered to obtain gain\cite{chiao4}.
However, spontaneous emission and the short excited state life time would cause such gain doublets
to be very difficult to sustain. It is important to have a steady state gain with a lifetime
longer than the pulse duration $\tau$ to avoid transient effects and the associated 
complications. 
Second, the medium must be transparent since opaque anomalous dispersion has been long known
and has resulted in controversies in terms of interpretations. While ideally the dispersion
shown in Fig.1 is transparent, residual absorption and the associated loss are often
present and cannot be simply discarded in experimental situations. Third, in order to 
show superluminal light pulse propagation in a linear regime, one must employ a very weak
light pulse for which the photon number is far less than the atomic number in order to avoid
Raman gain saturation. Fourth, in order to achieve a reasonable accuracy in the measurement, a system should
be designed to demonstrate a negative group velocity. In this case, the pulse advancement under 
conventional experimental situation will be substantially large compared with
commonly obtained accuracy (about 1ns). A number of other experimental conditions
also have to be considered such as atomic density, polarization decay time, etc. and they will be
discussed in the following sections as well. 

\subsection{Experimental Setup}

The experiment is performed using an atomic Cesium (Cs) vapor cell at 30$^{o}$C and the main setup
is shown in Fig.4. The cesium atoms are confined in a 6-cm-long Pyrex glass cell coated with
paraffin for the purpose of maintaining atomic ground-state spin polarization. The atomic cell is 
placed inside a magnetic shield made of a thin layer of high-$\mu$ metal material 
inside which the Earth
magnetic field is reduced to sub-milli-Gauss level. A Helmholtz coil (not shown in Fig.4) 
produces a uniform magnetic field inside the magnetic shield parallel to the propagation
direction of all optical beams. This uniform field is approximately 1 Gauss serving the
purpose of defining a quantization axis for optical pumping. Inside the magnetic shield,
the air temperature is controlled using a heater servo system in order to control the temperature of
the Cesium cell. During data acquisition, this control system is turned off to avoid any
stray magnetic field. Having good thermal insulation, the temperature of the atomic cell
remains the same during the data acquisition time.

In region-I of Fig.4, two optical pumping laser beams prepare almost all Cesium atoms into the
ground-state hyperfine magnetic sublevel $6S_{1/2},\,|F=4,m=-4\rangle$ that serves as the state
$|1\rangle$ in Fig.3. Laser-1 is a narrow linewidth diode laser locked to the 852-nm 
$D_{2}$ transition of Cs using a Lamb-dip technique and empties the $6S_{1/2},\,F=3$ 
hyperfine ground states\. Laser-2 is a broadband tunable Ti:sapphire laser tuned to 
the 894-nm $D_{1}$ transition of Cesium. The linewidth of laser-2 covers transitions from
both the $6S_{1/2},\,F=4$ and $F=3$ hyperfine ground states to the $6P_{1/2}$ excited
state. Both laser beams are initially linear polarized and are turned into left-hand
polarization ($\sigma-$) using a quarter-wave plate placed before the atomic cell. 
Inside the vapor cell, Cesium atoms collide with the paraffin-coated glass walls, the
atoms change their velocities inside the Doppler profile. However, their ground
state spin polarizations are not changed during collisions. Hence, all atoms
inside the entire Doppler broadening profile are optically pumped into the ground state
$|F=4,m=-4\rangle$ quickly. The mean dephasing time of the ground state spin polarization
of Cesium atoms in a paraffin coated cell is of order a fraction of a second. 

\begin{figure}
\hspace*{-.1 in}
\BoxedEPSF{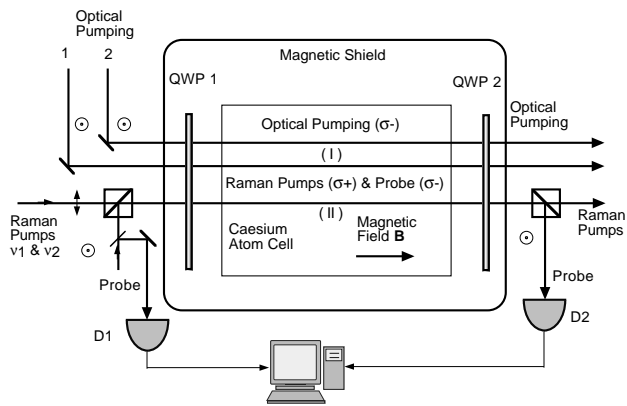 scaled 400}
\vspace*{.2 in}
\caption{Schematic experimental set up. 
Two optical pumping beams tuned to the Cesium (Cs) atomic D1 and D2 transitions prepare the
atoms in its ground state hyperfine sublevel $|F=4, m=-4\rangle$.
Two Raman pump beams and a Raman probe beam derived from a common narrow
linewidth diode laser propagate colinearly parallel to a small magnetic field B through the atomic cell. 
Two $\lambda/4$-plates
(QWP1 and 2) are used to prepare the three light beams into the corresponding circular polarization states
and then separate them for analysis.}
\end{figure}

In region-II, three light beams derived from the same narrow linewidth diode laser propagate
colinearly through the cell. Two strong continuous-wave (CW) Raman pump beams are right-hand
polarized ($\sigma+$) and a weak Raman probe beam is left-hand polarized ($\sigma-$). 
Using 3 acousto-optical modulators (AOM's), the frequency difference of the two Raman pump
beams can be tuned continuously over a few MHz while the probe beam can also be tuned in
frequency and can be operated in both CW or pulsed mode. The typical carrier frequency of the
AOM's is 80 MHz and the linewidth is about 20 kHz. A residual optical beam that is shifted
in frequency by 80 MHz generated from the same AOM which modulates the probe beam is also available
for the refractive index measurement. 

\subsection{Experimental Methods and Results}

First, we operate the Raman probe beam in a tunable CW mode to measure the gain 
and refractive index of the atomic system as a function of the probe frequency detuning. 
Fig.5 shows the measured gain coefficient and the refractive index. 
In order to obtain the gain coefficient, we first
measure the intensity of the transmitted probe beam as a function of probe frequency. 
We then extract the gain coefficient.

\begin{figure}
\hspace*{-.4 in}
\BoxedEPSF{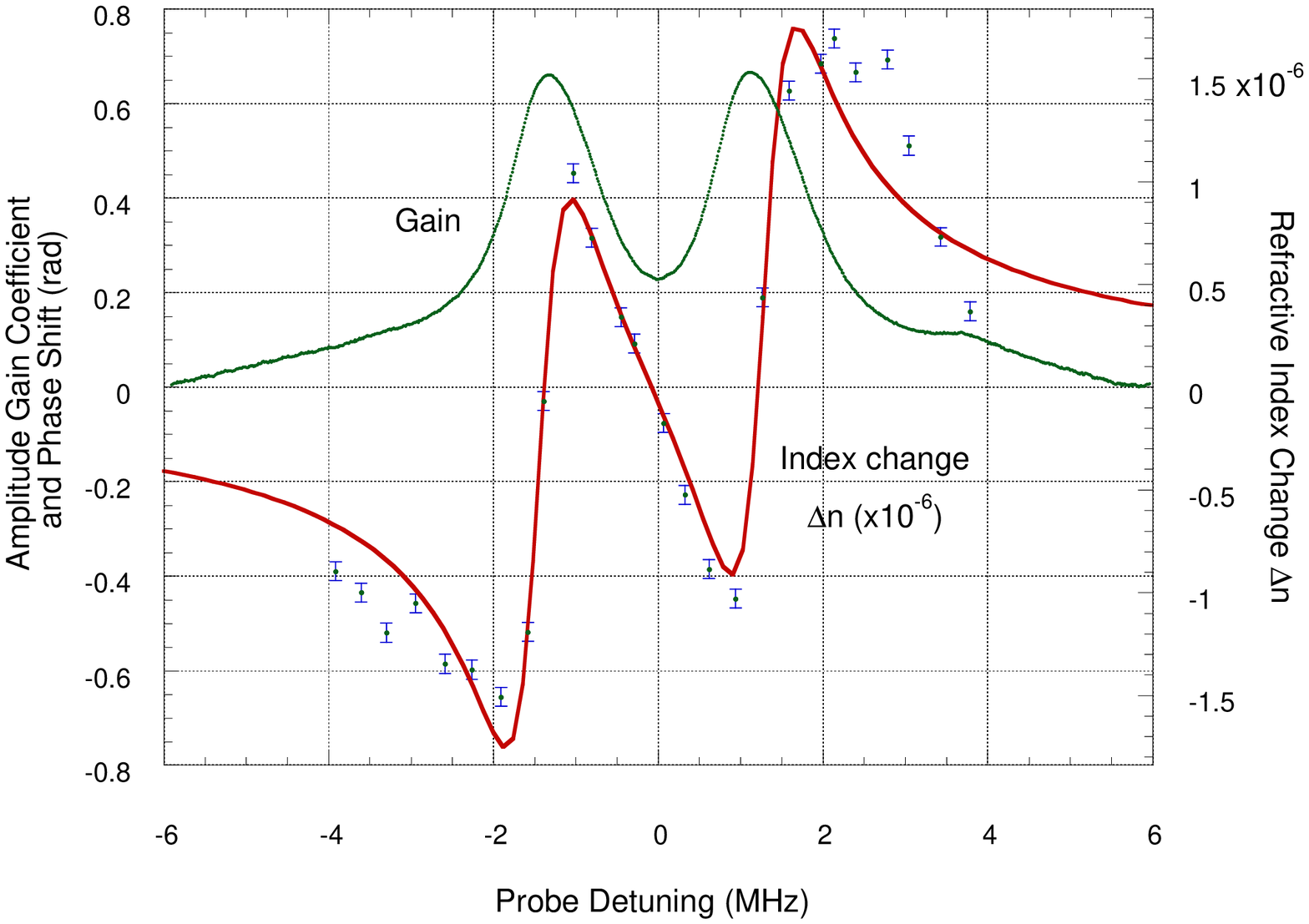 scaled 420}
\caption{Measured refractive index and gain coefficient. 
The superposed curve over the index data is obtained using Eq. (30)
with parameters $\nu_{0}, \Delta\nu$, and $\gamma$ obtained experimentally.
The length of the cell is 6 cm.}
\end{figure} 

The refractive index is measured using a radio-frequency (rf) interferometric technique. 
We send two CW light beams from the probe beam input port of the polarized beamsplitter. 
These two optical beams are different in frequency by $\Omega\approx$ 80 MHz 
(derived from the zeroth and the first order output of the probe AOM).
A weak probe beam is tuned in frequency to be close to the Raman pump
frequency and a strong local oscillator field shifted by $\Omega\approx$ 80 MHz 
is outside of the interesting frequency range.
The rf beating signal is detected using two fast detectors D1 and D2. 
We AC couple both output signals and record the beating signal of D2
\begin{equation}
V_{on}(t,\phi_{on}(\nu))=V_{0}\,cos[\Omega\,\delta t-\phi_{on}(\nu)-\phi_{0}]
\end{equation}
as a function of the AOM frequency $\Omega$.
using the AC coupled D1 voltage signal as the trigger.
We then apply a least-squares fitting procedure using Eq.(43)
to obtain the frequency dependent phase shift $\phi_{on}(\nu)$. In Eq.(43),
$\delta t$ is a residual difference between the electrical delays
of the trigger channel and the signal channel and is minimized. 
$\phi_{0}$ is a fixed phase factor. Of course, the fitting procedure 
always only yields the combined phase shift,
\begin{equation}
\phi_{on}(\nu)-\delta\Omega\,\delta t=2\pi n(\nu)\,L/\lambda -\delta\Omega\,\delta t
\end{equation}
where $\delta\Omega=\Omega-\Omega_{0}$ ($\Omega_{0}$=80 MHz)
is the change in frequency relative to the mean carrier frequency. In order to achieve
high accuracy of the refractive index measurement, one must separately measure
the factor $\delta\Omega\delta t$. To do this, we tune the diode laser outside of 
the Doppler profile where the atoms are irrelevant and Eq.(43) becomes:
\begin{equation}
V_{off}(t)=V_{1}\,cos[\Omega\,\delta t -\phi_{off}-\phi_{0}]
\end{equation}
since the phase shift here $\phi_{off}=0$. Hence, we directly measure the
factor $\delta\Omega\delta t$ and can subtract it out from the combined phase 
shift in Eq.(44) to obtain the the refractive phase shift to obtain the refractive index:
$\phi_{on}(\nu)=2\pi n(\nu)\,L/\lambda$. The result is shown in Fig.5.
The superimposed curve is obtained from Eq.(30) using parameters obtained from the gain measurement. 
From Fig.5, we readily see that a negative change of $\Delta$n= -1.8x10$^{-6}$ in the refractive 
index occurs over a narrow probe frequency range of $\Delta\nu$= 1.9 MHz between the two gain lines. 
Using the expression of the group-velocity index, we obtain the result 
$n_{g}=-330 (\pm 30)$ in that frequency region. The 10\% error reflects the accuracy of 
the phase measurement.

Next, a pulsed Raman probe beam is employed to observe the
superluminal propagation. A near Gaussian probe pulse with a 2.4 $\mu$sec FWHM is generated by applying a
biased sinusoidal electronic signal to the probe beam A/O modulator. The repetition rate is 50 kHz.
A portion of the pulsed probe beam is divided at a
beam-splitter before the atomic cell and aligned onto photodiode D1 as a reference. Because the total
number of atoms in the probe volume limits the maximum net energy gain of the probe pulse, we use a very
weak probe beam ($\approx 1 \mu$W) in order to avoid saturation and hence to optimize the anomalous dispersion. 
A high sensitivity avalanche photodiode, reverse-biased below breakdown, serves as detector D2 to
measure the weak probe pulse that propagates through the atomic cell. The photoelectric current produced
by detector D2 is converted to a voltage signal using a 500-$\Omega$ load resistor and recorded by a digitizing
oscilloscope using a synchronized output signal from the pulse generator as the trigger. Pulses from
detector D1 are also recorded.  

In order to measure the pulse propagation time, we first tune the diode
laser that produces the Raman pump and probe beams far off-resonance from the 852 nm Cesium D2 lines 
(by 2.5 GHz) to
measure the time-dependent probe-pulse intensity. When the laser is placed far off-resonance, the atoms
have no effect and the probe pulse propagates at the speed $c$ inside the cell. We then tune the diode laser
back to within the Doppler absorption profile and lock it on its side. Using the same synchronized pulse
generator output signal as the trigger, we record the time-dependent probe pulse intensity measured by
detector D2. We verify that no systematic drift is present by tuning the laser off-resonance again by the
same amount and record the probe pulse signal; the two off-resonance pulses are identical to within less
than 1 ns. Probe pulses both on and off-resonance are shown in Fig.6 (average of approximately 1000 pulses). 
\begin{figure}
\hspace*{-.2 in}
\vspace*{.1 in}
\BoxedEPSF{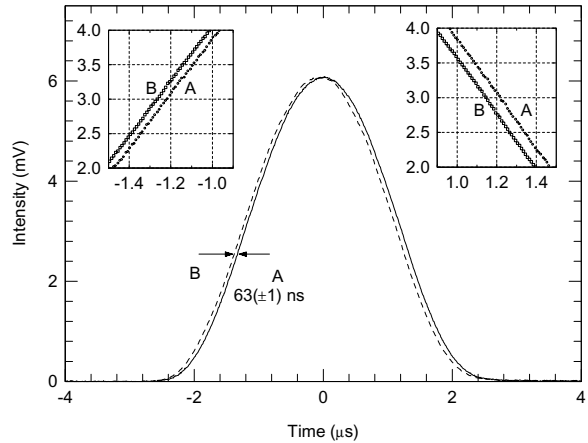 scaled 450}
\caption{Measured pulse advancement for a light pulse traversing through the 
Cesium vapor. (A), light pulse far off-resonance from the Cesium D2 transitions propagating 
at speed c through 6 cm of vacuum. (B), same
light pulse propagating through the same Cs-cell near resonance with a negative 
group velocity $-c/315$. Insets show the front
and trail parts of the pulses. A and B are both the average of 1000 pulses. 
Off-resonance pulse (A) is normalized to the
magnitude of (B).}
\end{figure}
Curve A in Fig.6 shows the light pulses far off-resonance from the Cesium D2 transitions propagating 
at speed c through 6 cm of vacuum. Curve B shows the same
light pulse propagating through the same Cs-cell near resonance at a negative 
group velocity $-c/315$. 
Probe pulses on resonance show a 40\%
transmittance and this is due to the broadband absorption of those atoms reverse pumped away from the  
state $|F=4, m=-4\rangle$. 
It is evident that there is almost no change in the pulse shape. The front edges and the trailing
edges of the pulses are shown in the insets; both edges are shifted forward by the same amount. 

Using a
least square fitting procedure, we obtain a pulse advancement shift of 63 ($\pm$1) nsec. Compared with the 0.2
nsec propagation time for light to traverse the 6-cm length of the atomic cell in vacuum, the 63 nsec
advancement gives an effective group index of $n_{g}= Ð315 (\pm 5)$. This is in good agreement with 
that inferred from the refractive index measurement. The pulses measured with detector
D1 are also recorded in the sequence of the off-, on-, off-resonance pulse propagation measurements and
are found to be identical to within 1.5 ns. 

We note here that the measured superluminal pulse propagation inside the transparent anomalous dispersion
medium is a linear effect. We further estimate the photon number per pulse and the interacting atomic number to
show that there is no saturation effect present. In the experiment, the measured voltage signal peak strength
is $V_{p}=\alpha\xi\,R G\, \dot{N}_{ph}\hbar\omega_{0}$, where $\alpha \approx 0.5 A/W$ is the photo responsivity
of the avalanche photo detector and $\xi=0.2$ is an effective efficiency of the detection imaging system.
 R=500 $\Omega$ is the load resistance. $G\approx 80$ is the avalanche gain.
$\hbar\omega_{0}=1.5 eV$ is the photon energy. Hence, we obtain the peak photon rate $\dot{N}_{ph}\approx 
5\times10^{12} /sec$. Using the $2.4\, \mu sec$ FWHM as the pulse duration, each probe pulse contains approximately
$1.2\times 10^{7}$ photons. On the other hand, inside the volume of the probe light beam $\pi w_{0}^{2}\,L$,
there are on average $N\,\pi w_{0}^{2}\,L$ atoms at any given moment, where $N\approx 10^{11} cm^{-3}$ is 
the atomic density. The beam radius is approximately $90\, \mu m$ and the Cesium cell is of a length 6 cm, the
atoms inside the beam volume is approximately $1.4\times10^8$ at any given moment.
However, since atoms are coming in and out of this volume within an average time
of $2\,w_{0}/V_{R}\approx 1\, \mu sec$, during the $2.4\, \mu sec$ pulse duration, there
are approximately $3.4\times 10^{8}$ atoms inside the volume of the light pulse, much larger
than the photon number per pulse. Hence, gain saturation effects are insignificant
and the observed superluminal pulse propagation is a linear effect.

To further analyze the linearity of the response of the anomalous dispersion medium, we compute the
pulse area of the Raman transition $\Omega_{R}\cdot\tau$. Here $\Omega_{R}=|\Omega_{1}\Omega_{2}/\Delta_{0}|$
is the effective Raman Rabi frequency.
By using Eq.(29) and by noting that 
$\Omega_{2}=(\mu_{02}\cdot\hat{e_{2}})E_{2}/2\hbar$, we obtain
\begin{equation}
M\,|E_{2}|^{2}=N\,\frac{\hbar}{\pi\varepsilon_{0}}\left|\frac{\Omega_{1}\Omega_{2}}{\Delta_{0}}\right|^{2},
\end{equation}
where $N$ is the atomic density. Furthermore, we have for the probe electric field $E_{2}$,
$P_{2}= 2\varepsilon_{0}cA|E_{2}|^{2}=\dot{N}_{ph}\,\hbar\omega_{0},$ where $P_{2}$ is the Raman probe power,
$A=\pi w_{0}^{2}$ is the beam cross-section, 
and $\hbar\omega_{0}$ is the energy per photon. Hence we obtain from Eq.(46)
$$
\left|\frac{\Omega_{1}\Omega_{2}}{\Delta_{0}}\right|^{2}=\frac{\pi M\omega_{0}\dot{N}_{ph}}{2NcA} \,\,. 
$$
We further recall from Eq.(28) that the peak Raman gain $G=gL$ at one of the resonances is $G=\pi\,ML/\gamma\lambda$.
Therefore, we obtain for the effective Raman Rabi frequency
\begin{equation}
\Omega_{R}=\left|\frac{\Omega_{1}\Omega_{2}}{\Delta_{0}}\right|=\sqrt{\frac{\pi G\gamma\dot{N}_{ph}}
{N\pi w_{0}^{2}L}}. 
\end{equation}
Using the experimental parameters of $\dot{N}_{ph}=5\times 10^{12}/sec$, $N\pi w_{0}^{2}L=1.4\times 10^{8}$,
peak gain $G=0.7$, and $\gamma=0.45 \,MHz$, we obtain $\Omega_{R}=1.25 \times 10^{5}/sec$. This is far smaller than
the typical Raman detuning $\Delta\nu\approx 1.4 \,MHz$. Recall that during the
$2.4 \,\mu sec$ pulse duration, the average dwell time for the atoms is $\Delta\tau=2 w_{0}/V_{R}= 1 \mu sec$. 
Hence, the equivalent pulse area becomes $\Omega_{R}\Delta\tau \approx 0.13$. It is apparent that
saturation of Raman gain is negligible.

Next, we perform experiments to test the parameter dependence given by Eq.(40). Namely, the pulse
advancement $|\Delta T|$ is linear in the parameter $M$ which depends linearly on the Raman pump
power $P_{R}$ and the atomic density $N$. First, the pulse advancement $\Delta T$ is measured as a 
function of the total Raman pump power of the two gain lines using the method described above.
Fig.7 shows the experimental result. 
In Fig.8, we show the measured
parameter dependence of pulse advancement $|\Delta T|$ as a function of atomic density. 
We measure the optical density of the atomic medium in an ancillary linear absorption experiment while 
varying the temperature
of the atomic cell. The optical density of the atomic cell is proportional to the atomic density $N$.

\begin{figure}
\BoxedEPSF{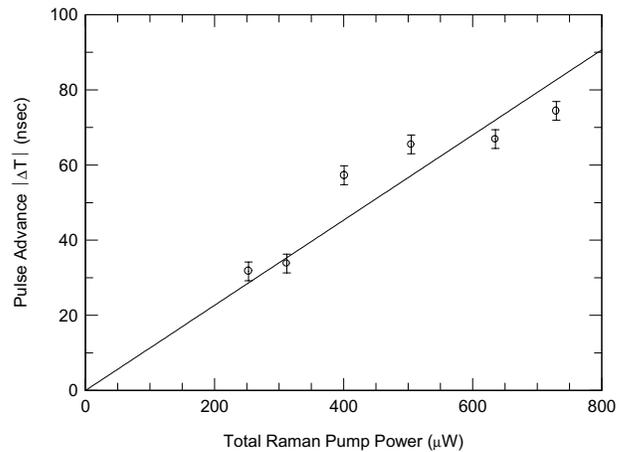 scaled 450}
\vspace*{.2 in}
\caption{Measured pulse advance $|\Delta T|$ dependence on total Raman pump power.
Solid line shows the least-square fitting based on the linear dependence shown in
Eq.(40)}
\end{figure}
\begin{figure}
\BoxedEPSF{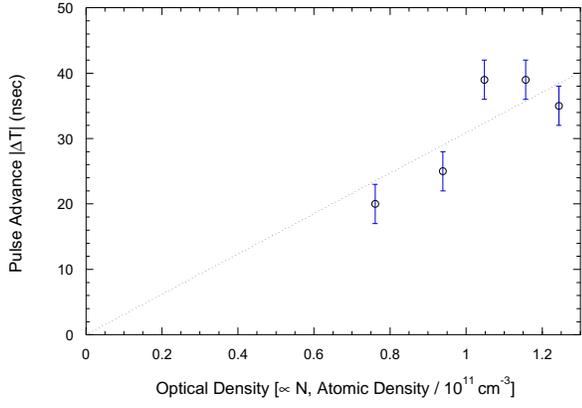 scaled 450}
\vspace*{.3 in}
\caption{Measured pulse advance $|\Delta T|$ dependence on the optical density of
the Cs chamber (linear in atomic density $N$).
Solid line shows the least-square fitting based on the linear dependence shown in
Eq.(40)}
\end{figure}

\begin{figure}
\BoxedEPSF{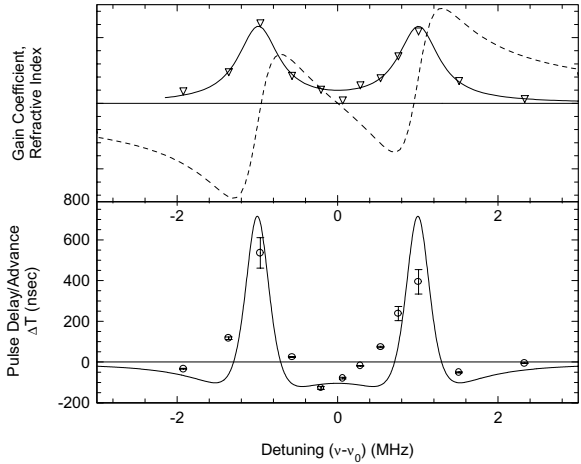 scaled 420}
\vspace*{.1 in}
\caption{Measured pulse gain coefficient and pulse delay/advance $\Delta T$ as a 
function of the carrier frequency of the light pulse. 
Dashed line shows the theoretical curve of refractive index using parameters extracted
from the gain coefficient using Eq.(30). Solid line shows the theoretical value of pulse
advance/delay based on the gain coefficient and refractive index.}
\end{figure}

Finally, we measure the pulse delay/advancement $\Delta T$ as a function of the pulse
carrier frequency $\nu$. Fig.9 shows the measurement results. For several values of the  
carrier frequency $\nu$,
the pulse gain is measured. We further record the pulse forms for the laser frequency tuned off-
and on- resonance. Using the measured pulse gain, we obtain the gain coefficient and fit the data
using Eq.(30). Using the fitting parameters, we compute the theoretical value of refractive index
and group delay/advancement plotted also in Fig.9. The large error bar for the pulse delay measurement
near the gain lines are due to the rapid change in group velocity index and gain, causing severe
pulse shape distortions. The simple model based on a Lorentz oscillator is in good agreement with
the experimental results.

\section{Discussion and conclusion}

As remarked by Lord Rayleigh\cite {rayleigh1}, the group velocity of a light pulse is the result of interference
between its various frequency components. Here we note that the measured negative and superluminal group velocity
of a light pulse propagating through a transparent anomalous dispersion medium is due to the physical effect of
``rephasing." Specifically, inside an anomalous dispersion medium, a longer wavelength (redder) component of a light
pulse has a slower phase velocity, contrary to the case of a normal dispersion medium. Conversely, a
shorter wavelength component (bluer) has a faster phase velocity. Inside a medium of refractive index $n$, the effective
wavelength of a light ray is modified: $\lambda'=\lambda\,/n$, where $\lambda$ is the vacuum wavelength.
Therefore, in a sufficiently strong anomalous dispersion medium, the 
redder incident ray will have a shorter wavelength and hence becomes a bluer ray, while an incident bluer ray
will have a longer wavelength to become a redder ray. This results in an unusual situation where the phases of the
different frequency components of a pulse become aligned at the exit surface of the medium earlier than even in the
case of the same pulse propagating through the same distance in a vacuum.

This highly unusual circumstance was poorly understood because it only occured in absorption lines previously.
Furthermore, inside an opaque medium, the majority of the incident light energy is 
absorbed and subsequently dissipated by the medium making it difficult to define the energy velocity of light. 
In the present experiment, an anomalous dispersion region is created in a transparent medium.
Hence it becomes possible to speak about the energy velocity
of a light pulse \cite{diener1,diener2}.

Here we  note that the physical mechanism that governs the observed superluminal light propagation 
has been traditionally viewed as ``virtual" reshaping \cite{chiao1,chiao7}. 
In the past, such superluminal light pulse propagation has been widely viewed as the
result of the amplification of the pulse's front edge and the absorption of its tail,
inspite it had been repeatedly pointed out that such reshaping is actually 
a ``virtual process" \cite{chiao1,chiao7}. 
In the present experiment, the 2.4$\mu\,$sec (FWHM)
probe pulse has only a 160 kHz bandwidth (FWHM) that is much narrower than the 
2.7 MHz separation of the two gain lines and the
probe pulse is placed in the middle of these gain lines spectrally. Hence, the probe 
pulse contains essentially no spectral
components that are resonant with the Raman gain lines to be amplified. Therefore, 
the argument that the probe pulse is advanced by amplification of its front edge does not apply. 
Furthermore, the average time an atom stays inside the volume of the Raman probe beam is 
$<1 \mu\,$sec, shorter than the 2.4$\mu$sec FWHM of the pulse. Hence, even if the atoms are
amplifying the probe pulse, both the leading and the trailing edges would be amplified as
both edges interact with atoms in the same state. This is not consistent with the experiment.
Hence it is worth emphasizing that the ``reshaping" of the
pulse can be only viewed as ``virtual reshaping" \cite{chiao1,chiao7} aimed at providing an intuitive understanding.
Strictly speaking, the superluminal light 
propagation observed here is the result only
of the anomalous dispersion region created with the assistance of two nearby Raman gain resonances. 
In other words, it is largely due to the rapid anomalous change of the refractive index as is shown in
Eq.(39) rather than the gain. When the gain coefficient becomes large, its effect appears as the
compression of the pulse, as indicated by Eq.(37) and (40).
We further stress that the
observed superluminal light propagation is a result of the wave nature of light \cite{born}. 
It can be understood using the classical theory
of wave propagation in an anomalous dispersion region where interference between different frequency components produces this
rather counterintuitive effect. 


\acknowledgments

{We thank R. A. Linke for several stimulating discussions. We thank J. A. Giordmaine, 
R. Y. Chiao, S. E. Harris, and E. S. Polzik for helpful discussions. We thank E. B. Alexandrov
and N. P. Bigelow for the use of the paraffin-coated Cesium cell.}


\end{multicols}
\end{document}